\documentclass[aps,prd,onecolumn,groupedaddress,showpacs,nofootinbib,amssymb]{revtex4}
\usepackage[dvips]{graphicx}
\usepackage{amssymb}
\usepackage{amsmath}
\usepackage{graphicx,,color}
\usepackage{amsfonts}
\usepackage{bm}
\usepackage{cancel}
\usepackage{comment}
\newcommand\be{\begin{equation}}
\newcommand\ee{\end{equation}}

\allowdisplaybreaks[4]

\begin{document}

\title{Geometric Inflation and Dark Energy with Axion $F(R)$ Gravity}
%\title{Inflation and Eschatological Scenarios with $f(R)$ Gravity and Axion Dark Matter}
\author{S.D. Odintsov,$^{1,2}$\,\thanks{odintsov@ieec.uab.es}
V.K. Oikonomou,$^{3,4,5}$\,\thanks{v.k.oikonomou1979@gmail.com}}
\affiliation{$^{1)}$ ICREA, Passeig Luis Companys, 23, 08010 Barcelona, Spain\\
$^{2)}$ Institute of Space Sciences (IEEC-CSIC) C. Can Magrans s/n,
08193 Barcelona, Spain\\
$^{3)}$Department of Physics, Aristotle University of Thessaloniki, Thessaloniki 54124, Greece\\
$^{4)}$ Laboratory for Theoretical Cosmology, Tomsk State
University of Control Systems
and Radioelectronics, 634050 Tomsk, Russia (TUSUR)\\
$^{5)}$ Tomsk State Pedagogical University, 634061 Tomsk, Russia }

\tolerance=5000

\begin{abstract}
We present a model of $F(R)$ gravity in the presence of a string
theory motivated misalignment axion like particle materialized in
terms of a canonical scalar field minimally coupled with gravity,
and we study the cosmological phenomenology of the model,
emphasizing mainly on the late-time era. The main result of the
paper is that inflation and the dark energy era may be realized in
a geometric way by an $F(R)$ gravity, while the axion is the dark
matter constituent of the Universe. The $F(R)$ gravity model
consists of an $R^2$ term, which as we show dominates the
evolution during the early time, thus producing a viable
inflationary phenomenology, and a power law term $\sim R^{\delta}$
with $\delta\ll 1 $ and positive, which eventually controls the
late-time era. The axion field remains frozen during the
inflationary era, which is an effect known for misalignment
axions, but as the Universe expands, the axion starts to
oscillate, and its energy density scales eventually as we show, as
$\rho_a\sim a^{-3}$.  After appropriately rewriting the
gravitational equations in terms of the redshift $z$, we study in
detail the late-time phenomenology of the model, and we compare
the results with the $\Lambda$CDM model and the latest Planck 2018
data. As we show, the model for small redshifts $0<z<5$ is
phenomenologically similar to the $\Lambda$CDM model, however at
large redshifts and deeply in the matter domination era, the
results are different from those of the $\Lambda$CDM model due to
the dark energy oscillations. For the late-time study we
investigate the behavior of several well-known statefinder
quantities, like the deceleration parameter, the jerk and $Om(z)$,
and we demonstrate that the statefinders which contain lower
derivatives of the Hubble rate have similar behavior for both the
$\Lambda$CDM and the axion $F(R)$ gravity model. We conclude that
the axion $F(R)$ gravity model can unify in a geometric way the
inflationary epoch with the dark energy era, and with the axion
being the main dark matter constituent.
\end{abstract}

%PACS numbers: 04.50.Kd, 95.36.+x, 98.80.-k, 98.80.Cq
\pacs{04.50.Kd, 95.36.+x, 98.80.-k, 98.80.Cq,11.25.-w}

\maketitle

\section{Introduction}

Currently the field of theoretical cosmology is challenged by
striking observational data, and solid answers must be found in
order to produce a viable cosmological description. We are living
in the era of precision cosmology, thus every theoretical model is
challenged and must be tested in numerous ways. The most important
unanswered for the moment observational and theoretical problems,
are related to the current accelerated expansion of the Universe
and to the dark matter issue. Specifically it is known for more
than twenty years that the Universe is expanding in an
accelerating way \cite{Riess:1998cb}, and recently it has been
verified by using different approaches, that the expansion rate
based on local data is different in comparison to the expansion
rate that the Universe had in the past, with the latter based on
the Cosmic Microwave Background anisotropy data
\cite{Aghanim:2018eyx}. This issue is currently known as the $H_0$
tension \cite{Riess:2011yx,Busti:2014dua}. There is also tension
in other null diagnostic quantities, this time related to Baryon
Acoustic Oscillations, like for example the quantity
$\mathcal{O}m(z)h^2$ (better stated the quantity $\Omega_M h^2$ at
$z\sim 2.34$) \cite{Sahni:2014ooa}, known as improved
$\mathcal{O}m(z)$ diagnostic. As it is stated in Ref.
\cite{Sahni:2014ooa}, this tension may alleviated by dynamically
evolving dark energy equation of state, or to our opinion,
modified gravity models with dynamical dark energy equation of
state may shed some light on these tensions. In fact, it is
possible that these tensions may be used as a test to verify the
possibility whether a modified gravity model is responsible for
the accelerating expansion of the Universe, and even discriminate
different modified gravity models that can produce such
phenomenological descriptions. For reviews on modified gravity and
dark energy see,
\cite{reviews1,reviews2,reviews3,reviews4,reviews5,reviews6} and
also Refs.
\cite{Capozziello:2002rd,Carroll:2003wy,Nojiri:2003ft,Nojiri:2007as,Nojiri:2007cq,Cognola:2007zu,Nojiri:2006gh}
for some streamline articles on the topic. Moreover, a
cosmographic approach to the dark energy issue may provide useful
insights for finding the exact dark energy equation of state
\cite{Benetti:2019gmo}.

On the other hand, the dark matter problem is also a fundamental
challenge for theoretical high energy physics and theoretical
cosmology for almost forty years. Continuous efforts seeking dark
matter particles with quite large masses, even of the order of
hundreds GeV or even TeV scale, had no results. There are several
candidates that could serve as the weakly interacting massive
particle (known as WIMPs) which we still seek, see for example
Refs. \cite{Oikonomou:2006mh}, however none of those has ever been
found, at least at up to present. The question is why to stick
with particle dark matter, while modified gravity can also
successfully describe certain aspects of particle dark matter? The
motivation to continue seeking for particle dark matter are the
observational data coming from galactic collisions, like the
Bullet cluster. Thus, there are two, last to our opinion, chances
for particle dark matter, ultralight stable particles, or
supersymmetry related particles that may be found by the Large
Hadron Collider in the far (or near) future. With regard to the
ultralight stable particles, the most important candidate
belonging to this category is the axion, or any string theory
motivated axion like particle
\cite{Marsh:2015xka,Sikivie:2006ni,Raffelt:2006cw,Linde:1991km}.
Currently there is a large number of researchers that study
several phenomenological implications of the axion, both in
astrophysics and cosmology and for an important stream of research
articles on this timely issue see for example
\cite{Marsh:2015xka,Marsh:2017yvc,Odintsov:2019mlf,Nojiri:2019nar,Nojiri:2019riz,Odintsov:2019evb,Cicoli:2019ulk,Fukunaga:2019unq,Caputo:2019joi,maxim,Chang:2018rso,Irastorza:2018dyq,Anastassopoulos:2017ftl,Sikivie:2014lha,Sikivie:2010bq,Sikivie:2009qn,Caputo:2019tms,Masaki:2019ggg,Soda:2017sce,Soda:2017dsu,Aoki:2017ehb,Masaki:2017aea,Aoki:2016kwl,Obata:2016xcr,Aoki:2016mtn,Ikeda:2019fvj,Arvanitaki:2019rax,Arvanitaki:2016qwi,Arvanitaki:2014wva,Arvanitaki:2014dfa,Sen:2018cjt,Cardoso:2018tly,Rosa:2017ury,Yoshino:2013ofa,Machado:2019xuc,Korochkin:2019qpe,Chou:2019enw,Chang:2019tvx,Crisosto:2019fcj,Choi:2019jwx,Kavic:2019cgk,Blas:2019qqp,Guerra:2019srj,Tenkanen:2019xzn,Huang:2019rmc,Croon:2019iuh,Day:2019bbh}
and references therein. Also there are currently many experiments
running and theoretical studies that may verify the existence of
$\mu$eV or even much smaller masses for the axion
\cite{Du:2018uak,Henning:2018ogd,Ouellet:2018beu,Safdi:2018oeu,Rozner:2019gba,Avignone:2018zpw,Caputo:2018vmy,Caputo:2018ljp,Lawson:2019brd}
with most of these experiments and theoretical proposals invoking
the axion photon conversion in a magnetic field
\cite{Balakin:2009rg,Balakin:2012up,Balakin:2014oya}. In addition,
the future LISA collaboration may reveal axionic effects coming
from gravitational waves related to superradiance of black holes
\cite{Arvanitaki:2019rax,Arvanitaki:2016qwi,Arvanitaki:2014wva}.
Notably, there is quite a number of works involving axions and
gravitational waves \cite{Satoh:2007gn}, mainly focusing on the
possibility of finding non-trivial polarized gravity waves. We
need to stress that several works in the literature, not related
to axions also discuss the circular polarization issue for
gravitational waves \cite{Inomata:2018rin,Kamionkowski:2000gb},
but axions involve Chern-Simons terms which may generate
inequivalent polarization modes \cite{Satoh:2007gn}. Another
interesting study is related to the effect of axions on the $H_0$
tension \cite{DEramo:2018vss}.

In view of the above major modern cosmological issues, in this
paper we shall study an $F(R)$ gravity model in the presence of a
misalignment axion canonical scalar field with the standard
(approximate) $V(\phi)\sim m_a^2\phi^2$ scalar potential, where
$m_a$ is the axion mass. The $F(R)$ gravity will contain the
standard Einstein-Hilbert term plus an $R^2$ term accompanied by a
positive non-integer power-law term $R^{\delta}$, with $\delta \ll
1$. The misalignment axion field has a primordial broken $U(1)$
Peccei-Quinn symmetry \cite{Dine:2004cq}, so during the whole
inflationary era it remains frozen in its primordial vacuum
expectation value, while it does not control the dynamical
evolution of the Universe at early times. On the other hand, as we
show, the dominant terms that drive the evolution at early times
are the standard Einstein-Hilbert term and the $R^2$ term. What we
achieve in this way is to obtain a viable inflationary era, and as
the Hubble rate value drops, the axion starts oscillations when
its mass is of the order $m_a\sim H$. By assuming that the axion
oscillates in a damped way, with the damping function being
slowly-varying, we demonstrate that the axion energy density
scales as $\rho_a\sim a^{-3}$, and its average equation of state
(EoS) parameter is $<w_a>=0$. Thus the axion behaves as dark
matter for all cosmic times that obey $m_a\gg H$. Then we turn our
focus on the late-time era and by introducing appropriate
variables, which by themselves are statefinders, we rewrite the
Friedmann equation in terms of the redshift $z$, and by choosing
physically motivated initial conditions we numerically solve the
Friedmann equation, focusing on redshifts in the range $z=[0,10]$,
thus covering the last stages of the matter domination era until
present time at $z=0$. As we show, the model can produce quite
interesting phenomenology, and the predicted cosmological
parameters match those of the $\Lambda$-Cold-Dark-Matter
($\Lambda$CDM) model, at least at present time. Also as we show,
deviations from the $\Lambda$CDM occur for larger redshifts, and
the deviations are enhanced, especially for cosmological
quantities such as the deceleration parameter and other
statefinder parameters which depend on higher derivatives of the
Hubble rate. This issue though depends on the initial conditions,
and also we discuss possible remedies that may soften these dark
energy oscillations during the matter domination era. As we
conclude, this model of $F(R)$ gravity is some sort of dynamically
evolving dark energy model and we briefly discuss future
perspectives of this work.

This paper is organized as follows: In section II we present the
$F(R)$ gravity axion model in some detail, and we discuss the
essential features of the misalignment axion model. In addition,
we show that the energy density of the axion scales as $a^{-3}$
and that the averaged axion EoS parameter is zero. Moreover, we
discuss the oscillating era of the axion in further detail and we
explicitly demonstrate the effects of the assumed slowly-varying
damped oscillation of the axion on the Friedmann equation. In
addition, we examine the phenomenology of the inflationary era for
the axion-$F(R)$ gravity model, and we quantify our claim that the
driving dominant terms are the Einstein-Hilbert term and the $R^2$
term. In section III we study in detail the late-time era and we
show that the axion-$F(R)$ gravity model produces a viable
late-time phenomenology. Also we discuss in some detail certain
aspects of the model, related to statefinder parameters that may
indicate whether a model of this type can actually be the correct
physical description of the Universe, and in addition we discuss
several other theoretical issues related to the model. Finally,
the conclusions of our study along with a discussion on the future
perspectives of this work follow in the end of the article.

\section{Description of the Model, Cosmological Dynamics and the Axion Scalar Evolution}

\subsection{The Misalignment Axion-$F(R)$ Gravity Model}

The axion-$F(R)$ gravity action we shall consider in this work,
has firstly appeared and briefly discussed at the end of our
previous work \cite{Odintsov:2019evb}, and it has the following
form,
\begin{equation}
\label{mainaction} \mathcal{S}=\int d^4x\sqrt{-g}\left[
\frac{1}{2\kappa^2}F(R)-\frac{1}{2}\partial^{\mu}\phi\partial_{\mu}\phi-V(\phi)+\mathcal{L}_m
\right]\, ,
\end{equation}
where $\kappa^2=\frac{1}{8\pi G}=\frac{1}{M_p^2}$, with $G$ being
Newton's gravitational constant and $M_p$ being the reduced Planck
mass. Also the Lagrangian $\mathcal{L}_m$ contains all the perfect
fluids that are assumed to be present in the theory. The form of
the $F(R)$ gravity which we shall choose is the following,
\begin{equation}\label{starobinsky}
F(R)=R+\frac{1}{M^2}R^2-\gamma \Lambda
\Big{(}\frac{R}{3m_s^2}\Big{)}^{\delta}\, ,
\end{equation}
and there is a very specific reason for choosing this $R^2$
corrected power law type of $F(R)$ gravity which we discuss in
later sections. Also $m_s$ in Eq. (\ref{starobinsky}) is
$m_s^2=\frac{\kappa^2\rho_m^{(0)}}{3}$. The parameter $\delta $ is
assumed to take positive values in the interval $0<\delta <1$, and
$\gamma$ is a dimensionless free parameter, while the parameter
$\Lambda $ is a parameter with mass dimensions $[m]^2$. The
parameter $M$ must be approximately $M= 1.5\times
10^{-5}\left(\frac{N}{50}\right)^{-1}M_p$ for early time
phenomenological reasons \cite{Appleby:2009uf}, with $N$ being the
$e$-foldings number. By assuming a flat Friedmann-Robertson-Walker
(FRW) geometric background of the form,
\begin{equation}
\label{metricfrw} ds^2 = - dt^2 + a(t)^2 \sum_{i=1,2,3}
\left(dx^i\right)^2\, ,
\end{equation}
upon varying the gravitational action with respect to the metric
and with respect to the scalar field, we obtain the following
gravitational equations of motion,
\begin{align}\label{eqnsofmkotion}
& 3 H^2F_R=\frac{RF_R-F}{2}-3H\dot{F}_R+\kappa^2\left(
\rho_r+\frac{1}{2}\dot{\phi}^2+V(\phi)\right)\, ,\\ \notag &
-2\dot{H}F=\kappa^2\dot{\phi}^2+\ddot{F}_R-H\dot{F}_R
+\frac{4\kappa^2}{3}\rho_r\, ,
\end{align}
\begin{equation}\label{scalareqnofmotion}
\ddot{\phi}+3H\dot{\phi}+V'(\phi)=0
\end{equation}
where $F_R=\frac{\partial F}{\partial R}$. In addition, the
``dot'' denotes differentiation with respect to the cosmic time
$t$, the ``prime'' denotes differentiation with respect to the
scalar field, while we also assumed that the only perfect fluid
present will be that of radiation, so $p_r=\frac{1}{3}\rho_r$. The
core assumption of this paper is that the axion scalar field
$\phi$ is the main constituent of cold-dark-matter, so we did not
include any other dark matter component in the action.

It is crucial to understand how the axion dynamics affects the
dynamics of the model at all eras. The axion model we shall
consider in this work is the misalignment axion model, so in the
next section we shall present in detail the dynamical evolution of
the axion from the inflationary era to the late-time eras. As we
already mentioned, this is a crucial component of our model, so we
discuss this issue in detail.

%%%%%%%%%%%%%%%%%%%%%%%%%%%%%%%%%%%%%%%%%%%%%%%%%%%%%%%%%%%%%%%%%%%%%%%%%%%%%%%%%%%%%%%%%%%%%%%%%%%%%%%%%%%%%%%%%%%%%%%%%%%%%%%%%%%

\subsection{The Misalignment Axion Field Dynamics: Inflation and Post-inflationary Eras Evolution}

As we already mentioned in the introduction, the axion scalar and
all axion like scalars, have attracted a lot of attention the last
years. The main reason for this is the absence of evidence for a
large mass WIMP, so currently the scientific community has focused
on mass scales of eV scale or quite smaller than that. The focus
in this paper will be on misalignment axion scalars, which result
from a primordial string theory motivated broken $U(1)$
Peccei-Quinn symmetry. The actual mechanism for the spontaneous
breaking of this primordial symmetry is not necessary for this
work, however, the important outcome is that during this
pre-inflationary epoch, the axion scalar obtains a large vacuum
expectation value, its mass is constant, and more importantly it
remains frozen in its vacuum expectation value. Essentially, it
contributes a cosmological constant in the Friedman equations, as
we will see. Eventually this freezing of the axion during the
inflationary era will enable the $F(R)$ gravity to control the
inflationary dynamics, and control even the subsequent reheating
era. Let us quantify the freezing of the axion like particle after
the primordial breaking of the $U(1)$ Peccei-Quinn symmetry, more
details on these issues can be found in \cite{Marsh:2015xka}. The
axion potential after the spontaneous breaking for the $U(1)$
Peccei-Quinn symmetry has the following approximate form at
leading order,
\begin{equation}\label{axionpotential}
V(\phi )\simeq \frac{1}{2}m_a^2\phi^2_i\, ,
\end{equation}
with $\phi_i$ being the (large) vacuum expectation value obtained
by the axion field after the breaking of the primordial $U(1)$
symmetry. In addition, as already mentioned, the axion field has a
nearly constant mass after the $U(1)$ breaking and for all the
subsequent eras, including the inflationary era.

During the inflationary era, the axion field is overdamped, and
this is quantified in the following initial conditions that
control its dynamics during inflation \cite{Marsh:2015xka},
\begin{equation}\label{axioninitialconditions}
\ddot{\phi}(t_i)\simeq 0,\,\,\,\dot{\phi}(t_i)\simeq
0,\,\,\,\phi(t_i)\equiv\phi_i=f_a\theta_a\, ,
\end{equation}
with $t_i$ being a time instance characterizing the inflationary
era. Also $f_a$ is an important constant of the axion particle
theory, called the axion decay constant, which plays a crucial
role in the axion phenomenology, and $\theta_a$ is the initial
misalignment angle. Hence during the inflationary era, the axion
contributes merely a cosmological constant in the gravitational
equations of motion (\ref{eqnsofmkotion}). This behavior continues
for all cosmic times for which $H\gg m_a$, however when the Hubble
rate drops significantly and becomes of the order $H\sim m_a$, the
axion field starts to oscillate in a damped way though. This can
be seen from the axion equation of motion, which is a canonical
scalar field equation of motion with potential
(\ref{axionpotential}),
\begin{equation}\label{scalarfieldeqnduringradiation}
\ddot{\phi}+3H\dot{\phi}+m_a^2\phi=0\, .
\end{equation}
As the Universe expands, the second term is merely a friction term
and the evolution is a damped oscillation, which commences
approximately when $H\sim m_a$ and continues until for all the
subsequent eras for which $m_a\gg H$. Let us quantify in detail
this axion oscillatory era because it is of crucial importance. We
shall assume that the axion oscillatory solution has the following
form \cite{Marsh:2015xka},
\begin{equation}\label{solutionaxionradandaft}
\phi (t)= \phi_i\mathcal{A}(t)\cos (m_a t)\, ,
\end{equation}
where $\phi_i$ is the initial value of the axion field after
inflation ends. The function $\mathcal{A}(t)$ is assumed to be a
slow-varying function, the dynamics of which are governed by the
following condition valid for all cosmic times for which
$m_a\succeq H$,
\begin{equation}\label{conditiononalpha}
\frac{\dot{\mathcal{A}}}{m_a}\sim \frac{H}{m_a}\simeq \epsilon \ll
1\, .
\end{equation}
Combining Eqs. (\ref{scalarfieldeqnduringradiation}) and
(\ref{solutionaxionradandaft}) and by keeping leading order terms
in the parameter $\epsilon$ we get,
\begin{equation}\label{scalarfieldeqnduringradiation1}
-\frac{2 \dot{\mathcal{A}} \sin (m_a t)}{m_a}-\frac{3 \mathcal{A}
H \sin (m_a t)}{m_a}=0\, ,
\end{equation}
so using  $H=\frac{\dot{a}}{a}$ we get the following approximate
differential equation,
\begin{equation}\label{newconditionforA}
\frac{d A}{A}=-\frac{3d a}{2a}\, ,
\end{equation}
with an analytic solution of the form,
\begin{equation}\label{solutionforA}
A\sim a^{-3/2}\, .
\end{equation}
The misalignment axion field is a canonical scalar field, so its
energy density and pressure are,
\begin{equation}\label{axionenergydensity}
\rho_a=\frac{\dot{\phi}^2}{2}+V(\phi)\, ,
\end{equation}
\begin{equation}\label{axionpressure}
P_a=\frac{\dot{\phi}^2}{2}-V(\phi)\, ,
\end{equation}
and the effective equation of state (EoS) parameter $w_a$ for the
axion scalar is $w_a=P_a/\rho_a$. Let us calculate these at
leading order in $\epsilon$, since these will be needed in the
sections that follow. Using the slow-varying oscillating solution
(\ref{solutionaxionradandaft}) the term $\frac{\dot{\phi}^2}{2}$
reads,
\begin{equation}\label{dotphianalytic1}
\frac{\dot{\phi}^2}{2}=\frac{m_a^2\phi_i^2}{2}\Big{[}\frac{\dot{A}}{m_a^2}\cos^2
(m_a t)+A^2\sin^2(m_at)-2A\frac{\dot{A}}{m_a}\cos (m_a t)\sin (m_a
t)\Big{]}\, ,
\end{equation}
and since $\frac{\dot{A}}{m_a}\sim \epsilon \ll 1$, the term
$\frac{\dot{\phi}^2}{2}$ can be approximated as follows,
\begin{equation}\label{dotphiterm}
\frac{\dot{\phi}^2}{2}\simeq \frac{1}{2}m_a^2\phi_i^2A^2\sin^2
(m_a t)\, .
\end{equation}
Moreover, the axion potential term is equal to,
\begin{equation}\label{axionpotentialterm}
V(\phi )=\frac{1}{2}m_a^2\phi_i^2A^2\cos^2 (m_a t)\, ,
\end{equation}
therefore by substituting Eqs. (\ref{dotphiterm}) and
(\ref{axionpotentialterm}) in the axion energy density
(\ref{axionenergydensity}), the later becomes,
\begin{equation}\label{axionenergydensityfinalpost}
\rho_a\simeq \frac{1}{2}m_a^2\phi_i^2A^2\, ,
\end{equation}
and in view of Eq. (\ref{solutionforA}), the axion energy density
reads,
\begin{equation}\label{axionenergydensityfinalpost}
\rho_a\simeq \rho_m^{(0)}a^{-3}\, ,
\end{equation}
where we have introduced the notation
$\rho_m^{(0)}=\frac{1}{2}m_a^2\phi_i^2$. Thus the axion energy
density scales as $\rho_a\sim a^{-3}$ for all cosmic times for
which $m_a\gg H$, in effect the axion scalar scales as a cold dark
matter perfect fluid. Since the axion mass is constant, the total
energy density of the cold dark matter scalar at present is
$\rho_m^{(0)}=\frac{1}{2}m_a^2\phi_i^2$, which indicates that its
magnitude is set by the initial conditions of the primordial axion
scalar, quantified in the initial vacuum expectation value of the
axion scalar $\phi_i$ which received it after the spontaneous
breaking of the primordial $U(1)$ Peccei-Quinn symmetry.

By substituting Eqs. (\ref{dotphiterm}) and
(\ref{axionpotentialterm}) in the pressure of the axion scalar
(\ref{axionpressure}), we obtain,
\begin{equation}\label{axionpressurefinal}
P_a\simeq \frac{1}{2}m_a^2\phi_i^2A^2 \Big{[}\sin^2 (m_a
t)-\cos^2(m_a t)\Big{]}\, ,
\end{equation}
and in effect, the EoS for the axion scalar $w_a=P_a/\rho_a$
reads,
\begin{equation}\label{axioneos}
w=\sin^2 (m_a t)-\cos^2(m_a t)\, ,
\end{equation}
which when integrated for an integration period yields $<w>=0$,
which also supports the fact that the axion scalar is a cold dark
matter particle.

Thus we showed that the axion scalar can act as a cold dark matter
particle, which we assumed that it constitute the whole dark
matter of the Universe, and we determined its dynamics during
inflation and for all the subsequent eras. Before closing, we
shall discuss all the essential phenomenological issues related
with the axion, in order to determine in a quantitative way the
order of magnitude of the axion scalar effects during the
inflationary era. We shall assume that the inflationary scale is
$H_I=10^{13}$GeV, and also that the current Hubble rate value
$H_0$ is that of the Planck observational data
\cite{Aghanim:2018eyx},
\begin{equation}\label{H0today}
H_0=67.4\pm 0.5 \frac{km}{sec\times Mpc}\, ,
\end{equation}
so $H_0=67.4km/sec/Mpc$ which is $H_0=1.37187\times 10^{-33}$eV,
hence $h\simeq 0.67$. In addition, the latest Planck data indicate
that the dark matter density $\Omega_c h^2$ are,
\begin{equation}\label{codrdarkmatter}
\Omega_c h^2=0.12\pm 0.001\, ,
\end{equation}
in which case, for $\phi_i=\mathcal{O}(10^{15})$GeV, the axion
mass compatible with the constraint (\ref{codrdarkmatter}) is of
the order $m_a\simeq \mathcal{O}(10^{-14})$eV. In the next
sections we shall investigate the inflationary and the late-time
phenomenology of the axion $F(R)$ gravity model, and as we
demonstrate it is possible to describe both the early and the
late-time acceleration in a geometric way, with the axion playing
the role of the cold dark matter particle.

\subsection{The Inflationary Era: $R^2$ Gravity Prevails}

In the previous section we demonstrated that the axion scalar
during the inflationary era contributes merely a cosmological
constant in the equations of motion of Eq. (\ref{eqnsofmkotion})
and in this section we shall calculate the order of magnitude of
this contribution and compare it to the $F(R)$ gravity terms
during the inflationary era. In this way we shall show that the
$R^2$ gravity terms control the dynamical evolution during the
inflationary era. Also we shall appropriately choose the values of
the free parameters and we shall show that the $F(R)$ gravity of
Eq. (\ref{starobinsky}) satisfies the $F(R)$ gravity viability
criteria.

We start off with the equations of motion (\ref{eqnsofmkotion}),
and specifically the Friedmann equation, which for the $F(R)$
gravity of Eq. (\ref{starobinsky}) and for the potential
(\ref{axionpotential}) reads,
\begin{equation}\label{friedmanequationinflation}
3H^2\left(1+\frac{2}{M^2}R-\delta \gamma
\Big{(}\frac{R}{3m_s^2}\Big{)}^{\delta-1}\right)=\frac{R^2}{2M}+(\gamma-\gamma
\delta
)\frac{\Big{(}\frac{R}{3m_s^2}\Big{)}^{\delta}}{2}-3H\dot{R}\Big{(}\frac{2}{M^2}-\gamma
\delta (\delta-1)\Big{(}\frac{R}{3m_s^2}\Big{)}^{\delta-2}\Big{)}+
\kappa^2\Big{(}\rho_r+\frac{1}{2}\kappa^2\dot{\phi_i}^2+\frac{1}{2}m_a^2\phi_i^2
\Big{)}\, .
\end{equation}
Now we shall choose the values free parameters $M$, $\gamma $ and
$\delta $ and we shall compare the order of magnitude of the terms
appearing in Eq. (\ref{friedmanequationinflation}) in order  to
see which terms drive the dynamical evolution of the Universe
during the inflationary era. The choice of the free parameters
$\gamma $ and $\delta $ is determined by the late-time era as we
see in the later section, so by choosing,
\begin{equation}\label{gammaanddelta}
\gamma=\frac{1}{0.5},\,\,\, \delta=\frac{1}{100}\, ,
\end{equation}
and also we assume that $\Lambda\simeq 11.895\times
10^{-67}$eV$^2$. As we shall see in the next section, these values
for $\gamma$ and $\delta$ can yield a very interesting late-time
evolution. Also $m_s$ was defined below Eq. (\ref{starobinsky}) so
$m_s^2\simeq 1.87101\times 10^{-67}$eV$^2$. Finally, the parameter
$M$ related to the $R^2$ term in Eq. (\ref{starobinsky}) for
phenomenological reasons must be chosen $M= 1.5\times
10^{-5}\left(\frac{N}{50}\right)^{-1}M_p$ \cite{Appleby:2009uf},
so for $N\sim 60$, $M$ is approximately $M\simeq 3.04375\times
10^{22}$eV.  Furthermore we assume that during the inflationary
era $\dot{H}\ll H^2$, and in effect the curvature is approximately
$R\simeq 12 H^2$, so for $H= H_I\sim 10^{13}$GeV, the curvature
scalar is approximately $R\sim 1.2\times 10^{45}$eV$^2$. Now let
us proceed in the comparison of the terms appearing in Eq.
(\ref{friedmanequationinflation}), and we start off by eliminating
the radiation term $\kappa^2\rho_r \sim e^{-N}$ which could be
eliminated from the beginning, since it does not affect the
evolution during inflation. The term $\dot{\phi}_i^2$ can also be
eliminated since the axion obeys the initial conditions
(\ref{axioninitialconditions}), so is frozen during the
inflationary era. Also, the values of $\phi_i$
 and $m_a$ were chosen in the previous section as
 $\phi_i=\mathcal{O}(10^{15})$GeV and $m_a\simeq
 \mathcal{O}(10^{-14})$eV, therefore the potential term is of the order $\kappa^2V(\phi_i)\sim \mathcal{O}(8.41897\times
 10^{-36})$eV$^{2}$, since $\kappa^2=1/M_p^2$ where $M_p$ is the reduced Planck mass $M_p\simeq 2.435 \times 10^{27}$eV. Let us now proceed to the curvature related
 terms, so the terms $R\sim 1.2\times \mathcal{O}(10^{45})$eV$^2$, also $R^2/M^2\sim \mathcal{O}(1.55\times
 10^{45})$eV$^2$. Finally the term $\sim \Big{(}\frac{R}{3m_s^2}\Big{)}^{\delta}\sim
 \mathcal{O}(10)$, also $\sim \Big{(}\frac{R}{3m_s^2}\Big{)}^{\delta-1}\sim
 \mathcal{O}(10^{-111})$ and lastly $\sim \Big{(}\frac{R}{3m_s^2}\Big{)}^{\delta-2}\sim
 \mathcal{O}(10^{-223})$. Clearly, the only dominant terms are
 those corresponding to the positive powers of the curvature,
 hence the Friedman equation (\ref{friedmanequationinflation}) at
 leading order during inflation is identical to the one
 corresponding to the vacuum $R^2$ model, that is,
\begin{equation}\label{friedmanequationinflationaux}
3H^2\left(1+\frac{2}{M^2}R\right)=\frac{R^2}{2M}-\frac{6H\dot{R}}{M^2}\,
,
\end{equation}
which can be rewritten,
\begin{equation}\label{patsunappendixinflation}
3\ddot{H}-3\frac{\dot{H}^2}{H}+\frac{2M^2H}{6}=-9H\dot{H}\, ,
\end{equation}
which can be solved by using the slow-roll assumption $\dot{H}\ll
H^2$ and it yields an approximate quasi-de Sitter evolution,
\begin{equation}\label{quasidesitter}
H(t)=H_0-\frac{M^2}{36} t\, .
\end{equation}
The phenomenology of the Jordan frame vacuum $R^2$ model with the
quasi-de Sitter evolution produces a viable inflationary era,
compatible with the latest Planck data \cite{Aghanim:2018eyx},
since the spectral index as a function of the $e$-foldings number
is $n_s\sim 1-\frac{2}{N}$ and the predicted tensor-to-scalar
ratio is $r\sim \frac{12}{N^2}$.

%%%%%%%%%%%%%%%%%%%%%%%%%%%%%%%%%%%%%%%%%%%%%%%%%%%%%%%%%%%%%%%%%%%%%VIABILITY CRITERIA OF FR

Thus in this section we demonstrated that the vacuum $R^2$ gravity
controls the evolution of the axion $F(R)$ gravity model, due to
the fact that the axion is dynamically frozen during inflation.
However as the Universe expands, when $m_a\succeq H$, the axion
starts to oscillate and behaves dynamically as cold dark matter,
as we showed in the previous section. In effect, at late-times it
behaves as a cold dark matter fluid the energy density of which
scales as $\rho_a\sim a^{-3}$. In the next section we shall
discuss the late-time phenomenology of the axion $F(R)$ gravity
model.

\section{Late-time Evolution and Cosmological Parameters}

The gravitational equations of motion (\ref{eqnsofmkotion}) and
(\ref{scalareqnofmotion}) can be written in a form similar to the
Einstein gravity case for a flat FRW spacetime,
\begin{align}\label{flat}
& 3H^2=\kappa^2\rho_{tot}\, ,\\ \notag &
-2\dot{H}=\kappa^2(\rho_{tot}+P_{tot})\, ,
\end{align}
where $\rho_{tot}=\rho_{\phi}+\rho_G+\rho_r$ is the total energy
density of the cosmological fluid and $P_{tot}=P_r+P_{a}+P_{G}$ is
the total pressure. In the case at hand, the total fluid consists
from the radiation perfect fluid with energy density $\rho_r$, the
axion scalar field fluid with energy density $\rho_{a}$, which is
given in Eq. (\ref{axionenergydensity}) and the geometric fluid
$\rho_{G}$ which at late-times will play the role of dark energy
and it is equal to,
\begin{equation}\label{degeometricfluid}
\rho_{G}=\frac{F_R R-F}{2}+3H^2(1-F_R)-3H\dot{F}_R\, .
\end{equation}
Accordingly the pressures can consist of the radiation, scalar
field and geometric part, with the pressure for the radiation
being $P_r=\frac{1}{3}\rho_r$, the pressure for the scalar fluid
being defined in Eq. (\ref{axionpressure}), and the pressure of
the geometric fluid being equal to,
\begin{equation}\label{pressuregeometry}
P_G=\ddot{F}_R-H\dot{F}_R+2\dot{H}(F_R-1)-\rho_G\, .
\end{equation}
All the fluids in the way we chose the respective energy momentum
tensors, do not interact between them and satisfy the continuity
equations,
\begin{align}\label{fluidcontinuityequations}
& \dot{\rho}_a+3H(\rho_a+P_a)=0\, , \\ \notag &
\dot{\rho}_r+3H(\rho_r+P_r)=0\, , \\ \notag &
\dot{\rho}_G+3H(\rho_G+P_G)=0\, .
\end{align}
The basic principle we would like to point out here is that the
geometric fluid controls apart from the early-time era, the
late-time acceleration era too, and the axion acts as a cold dark
matter dust.

For the late-time study, we shall express the Friedmann equation
in terms of the redshift $z$ which is defined as follows,
\begin{equation}\label{redshift}
1+z=\frac{1}{a}\, ,
\end{equation}
where we assumed that the present time scale factor, which
corresponds to $z=0$ is equal to one. Also, we shall introduce the
function $y_H(z)$ to quantify our study
\cite{Hu:2007nk,Bamba:2012qi}, which is defined as follows,
\begin{equation}\label{yHdefinition}
y_H(z)=\frac{\rho_{G}}{\rho^{(0)}_m}\, ,
\end{equation}
where $\rho^{(0)}_m$ is the present time energy density of cold
dark matter. In terms of the first Friedman equation (\ref{flat}),
the function $y_H(z)$ is written as,
\begin{equation}\label{yhfunctionanalyticzero}
y_H(z)=\frac{3H^2}{\kappa^2\rho^{(0)}_m}-\frac{\dot{\phi}^2}{2\rho^{(0)}_m}-\frac{V(\phi)}{\rho^{(0)}_m}-\frac{\rho_r}{\rho^{(0)}_m}\,
.
\end{equation}
The radiation energy density scales as
$\rho_r=\rho_r^{(0)}a^{-4}$, where $\rho_r^{(0)}$ is the present
time value of the radiation energy density, so
$\frac{\rho_r}{\rho^{(0)}_m}=\chi (1+z)^4$, where
$\chi=\frac{\rho^{(0)}_r}{\rho^{(0)}_m}\simeq 3.1\times 10^{-4}$.
The most interesting part is the scalar field part, so the
$\phi$-dependent terms in Eq. (\ref{yhfunctionanalyticzero}). At
late times, the axion field oscillates with a frequency $m_a$ as
we showed earlier, and by combining Eqs. (\ref{dotphiterm}) and
(\ref{axionpotentialterm}), the two terms in Eq.
(\ref{yhfunctionanalyticzero}) are equal to,
\begin{align}\label{endiamesieqn}
&
\frac{1}{\rho^{(0)}_m}\left(-\frac{\dot{\phi}^2}{2}-V(\phi)\right)
\\ \notag &
=-\frac{\phi_i^2m_a^2}{2\rho^{(0)}_m}\left(
\frac{\phi_i^2\dot{A}^2}{m_a^2}\cos^2(m_a t)
+A^2\phi_i^2\sin^2(m_a t)-\frac{2\dot{A}\phi_i^2}{m_a}A\cos (m_a
t)\sin (m_a t)+A^2\phi_i^2\cos^2(m_a t)\right)=-a^3\, ,
\end{align}
where we used Eq. (\ref{axionenergydensityfinalpost}) and the
definition $\rho_m^{(0)}=\frac{1}{2}m_a^2\phi_i^2$ we gave
earlier. Thus, in view of Eqs. (\ref{endiamesieqn}), and
substituting $\frac{\rho_r}{\rho^{(0)}_m}=\chi (1+z)^4$, the
function $y_H(z)$ of Eq. (\ref{yhfunctionanalyticzero}) is finally
written,
\begin{equation}\label{finalexpressionyHz}
y_H(z)=\frac{H^2}{m_s^2}-(1+z)^{3}-\chi (1+z)^4\, .
\end{equation}
where the parameter
$m_s^2=\frac{\kappa^2\rho^{(0)}_m}{3}=H_0\Omega_c=1.37201\times
10^{-67}$eV$^2$ was defined below Eq. (\ref{starobinsky}). Now let
us express the cosmological equation as a function of the variable
$y_H(z)$, so it can be shown that this is written as follows
\cite{Bamba:2012qi},
\begin{equation}\label{differentialequationmain}
\frac{d^2y_H(z)}{d z^2}+J_1\frac{d y_H(z)}{d z}+J_2y_H(z)+J_3=0\,
,
\end{equation}
where the functions $J_1$, $J_2$ and $J_3$ are defined as follows,
\begin{align}\label{diffequation}
& J_1=\frac{1}{z+1}\left(
-3-\frac{1-F_R}{\left(y_H(z)+(z+1)^3+\chi (1+z)^4\right) 6
m_s^2F_{RR}} \right)\, , \\ \notag & J_2=\frac{1}{(z+1)^2}\left(
\frac{2-F_R}{\left(y_H(z)+(z+1)^3+\chi (1+z)^4\right) 3
m_s^2F_{RR}} \right)\, ,\\ \notag & J_3=-3(z+1)-\frac{\left(1-F_R
\right)\Big{(}(z+1)^3+2\chi (1+z)^4
\Big{)}+\frac{R-F}{3m_s^2}}{(1+z)^2\Big{(}y_H(z)+(1+z)^3+\chi(1+z)^4\Big{)}6m_s^2F_{RR}}\,
,
\end{align}
where $F_{RR}=\frac{\partial^2 F}{\partial R^2}$. The above
differential equation must be solved by using appropriate initial
conditions, for a range of suitable redshift values that describe
the last stage of the matter domination epoch and the late-time
era up to present day. We shall focus on the interval
$z=[z_i,z_f]$ with $z_i=0$ and $z_f=10$, so the initial conditions
on the function $y_H(z)$ and its derivative are determined by the
last stages of the matter domination era. The Ricci scalar in
terms of the function $y_H(z)$ is written as follows,
\begin{equation}\label{ricciscalarasfunctionofz}
R(z)=3m_s^2\left( 4y_H(z)-(z+1)\frac{d y_H(z)}{d
z}+(z+1)^3\right)\, .
\end{equation}
Now an important issue is the initial conditions, and how these
affect the late-time phenomenology. Actually, as we will see, the
right choice of initial conditions may provide a reasonable
phenomenological picture for statefinder parameters that contain
higher derivatives of the Hubble rate. We shall consider the
following general choice of initial conditions for the redshift
$z_f=10$,
\begin{equation}\label{generalinitialconditions}
y_H(z_f)=\frac{\Lambda}{3m_s^2}\left(
1+\tilde{\gamma}(1+z_f)\right)\, , \,\,\,\frac{d y_H(z)}{d
z}\Big{|}_{z=z_f}=\tilde{\gamma}\frac{\Lambda}{3m_s^2}\, ,
\end{equation}
with the dimensionless parameter $\tilde{\gamma}$ will be assumed
to be $\tilde{\gamma}=\frac{1}{10^3}$ but in principle can take
larger values. We need to note that this parameter strongly
affects the large redshift behavior of the function $y_H$ and also
of the corresponding statefinder parameters, so eventually it may
affect the high redshift phenomenology and the dark energy
oscillations, which is always an issue in dynamical dark energy
modified gravity models. We shall take that as a free parameter,
and among other issues, we shall examine the effect of
$\tilde{\gamma}$ on the phenomenology of the model. The actual
determination of the initial conditions is always going to be some
variant form of the above, and perhaps some cosmographic approach
\cite{Benetti:2019gmo} may provide some more accurate form for
these. We shall assume that these are of the form
(\ref{generalinitialconditions}), so one of the main aims of this
section is also to investigate the effect of the initial
conditions on the cosmological parameters.

Now we can start making comparisons with the $\Lambda$CDM model,
in which case the Hubble rate is equal to,
\begin{equation}\label{lambdacdmhubblerate}
H_{\Lambda}(z)=H_0\sqrt{\Omega_{\Lambda}+\Omega_M(z+1)^3+\Omega_r(1+z)^4}\,
,
\end{equation}
where $H_0$ is the present day value of the Hubble rate which is
$H_0\simeq 1.37187\times 10^{-33}$eV according to the latest
Planck data \cite{Aghanim:2018eyx}, $\Omega_{\Lambda}\simeq
0.681369$ and $\Omega_M\sim 0.3153$ \cite{Aghanim:2018eyx}, while
$\Omega_r/\Omega_M\simeq \chi$, with $\chi$ being defined below
Eq. (\ref{yhfunctionanalyticzero}).

\begin{figure}[h!]
\centering
\includegraphics[width=18pc]{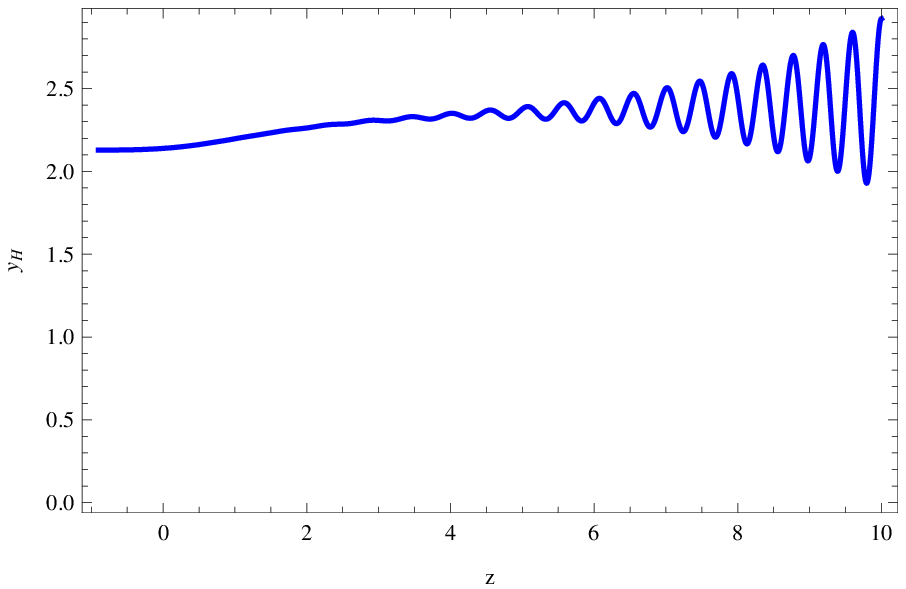}
\includegraphics[width=18pc]{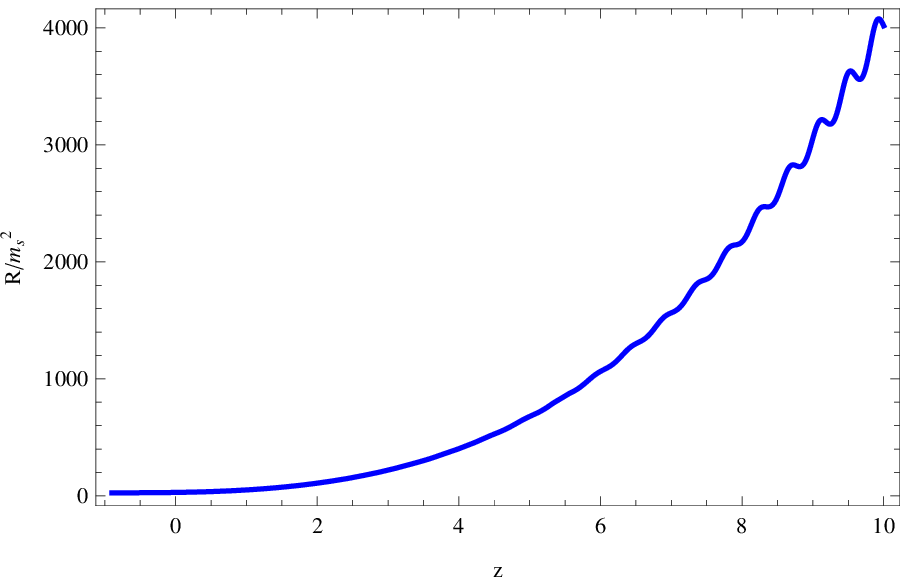}
\caption{The statefinder function $y_H$ for geometric dark energy
as a function of the redshift (left plot) and the scalar curvature
(right plot) for $\tilde{\gamma}=1/10^3$.} \label{plot1}
\end{figure}

\begin{figure}[h!]
\centering
\includegraphics[width=18pc]{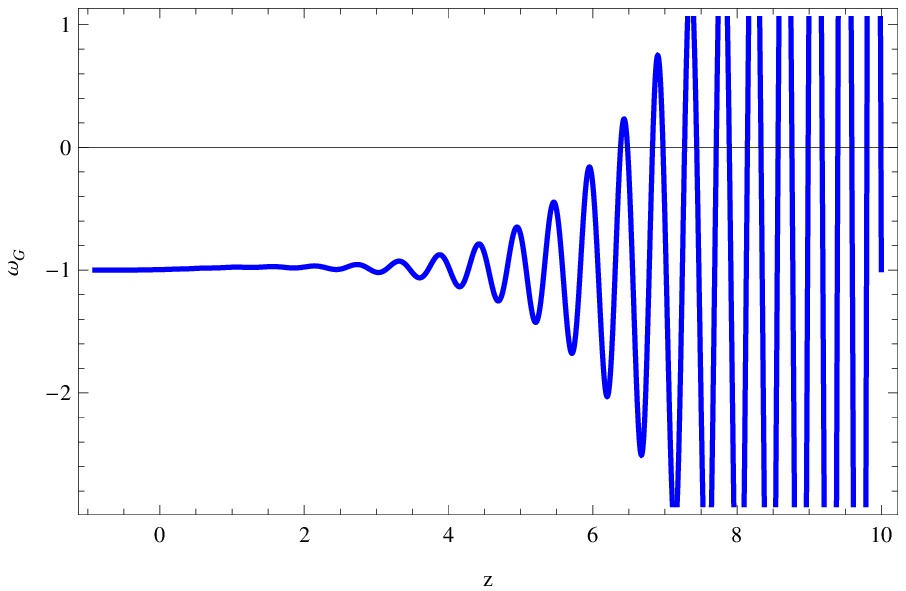}
\includegraphics[width=18pc]{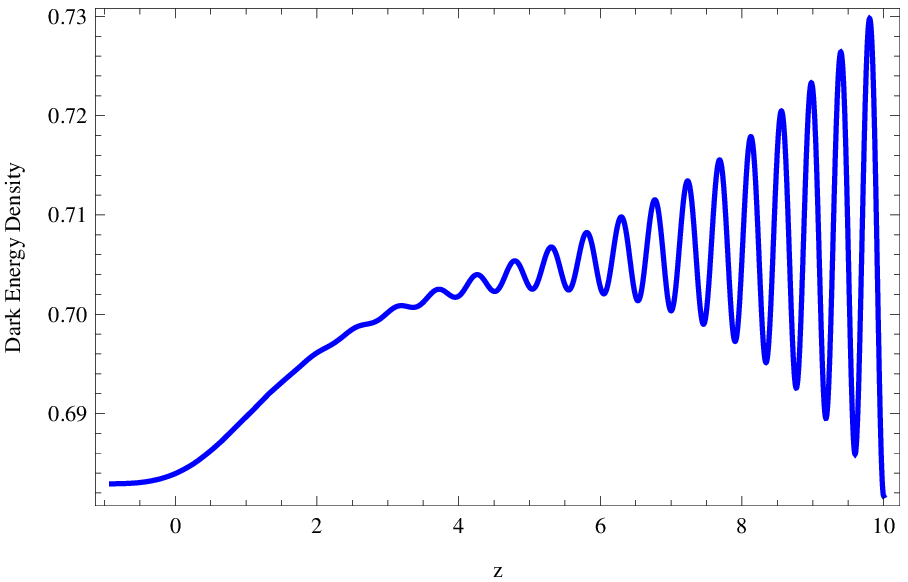}
\caption{The dark energy EoS parameter $\omega_G(z)$ (left plot)
and the dark energy density parameter $\Omega_G(z)$ (right plot)
as functions of the redshift for $\tilde{\gamma}=1/10^3$.}
\label{plot2}
\end{figure}

\begin{figure}[h!]
\centering
\includegraphics[width=18pc]{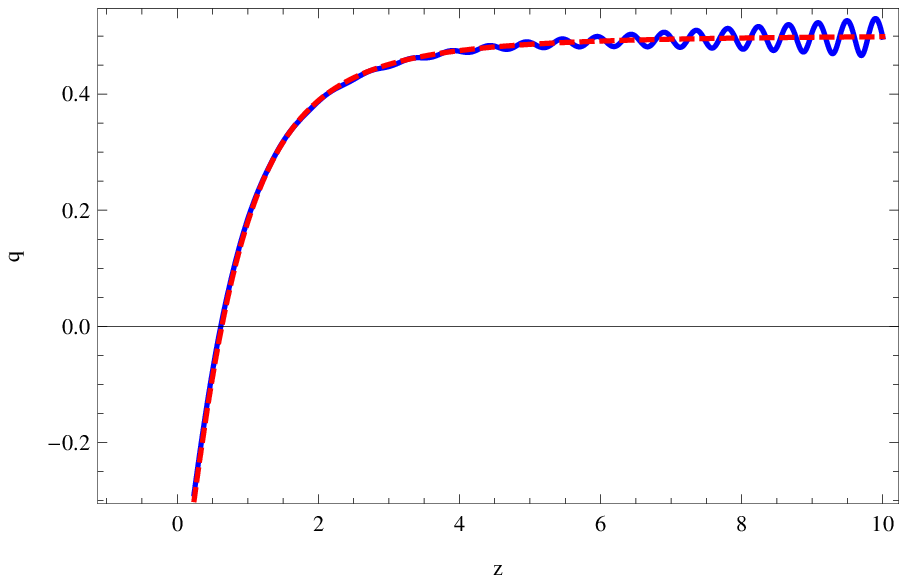}
\includegraphics[width=18pc]{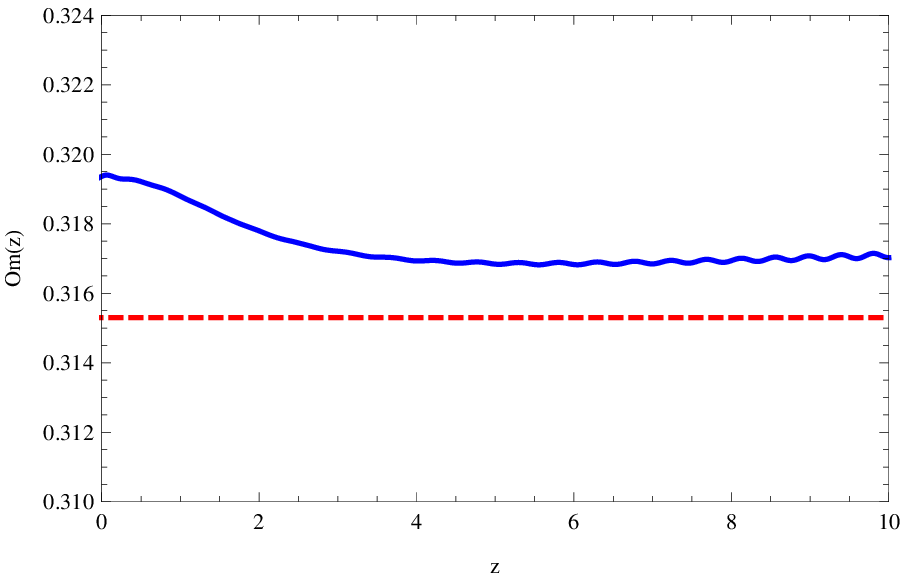}
\includegraphics[width=18pc]{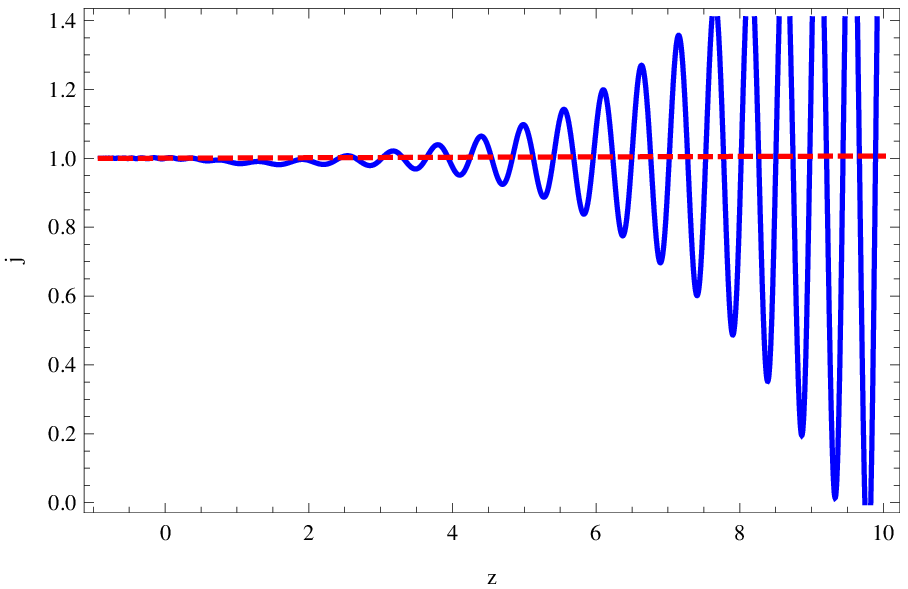}
\includegraphics[width=18pc]{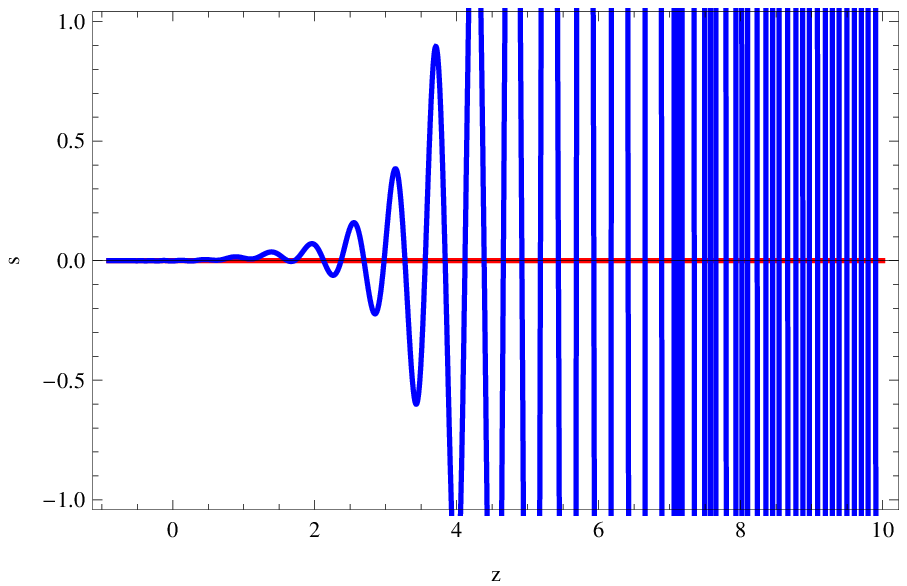}
\caption{The deceleration parameter $q$ (upper left plot), the
statefinder $Om(z)$ (upper right plot), the statefinder $j$ (lower
left) and the statefinder $s$ (lower right) as functions of the
redshift, for the axion $F(R)$ gravity model (blue curves) and for
the $\Lambda$CDM model (red curves), for $\tilde{\gamma}=1/10^3$.}
\label{plot3}
\end{figure}

\begin{figure}[h!]
\centering
\includegraphics[width=18pc]{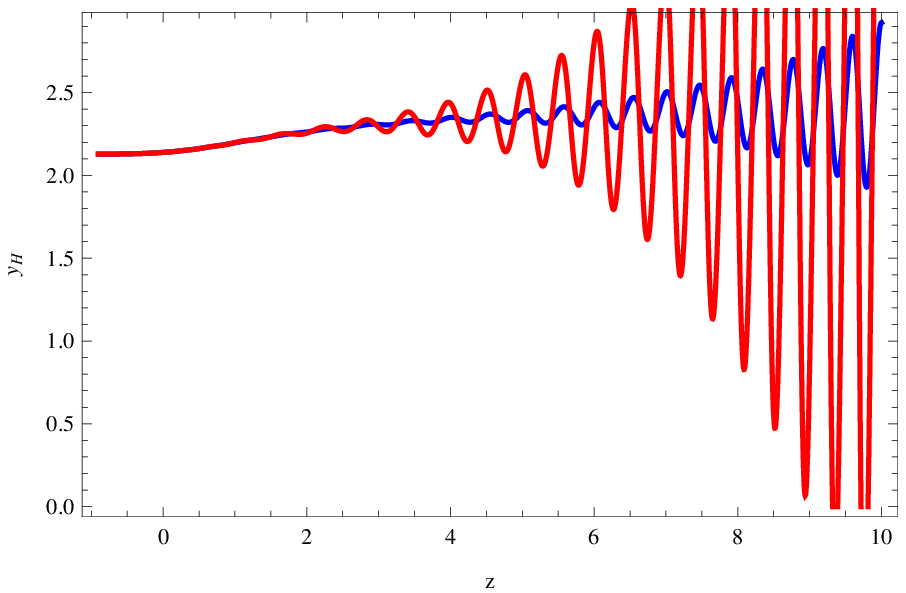}
\includegraphics[width=18pc]{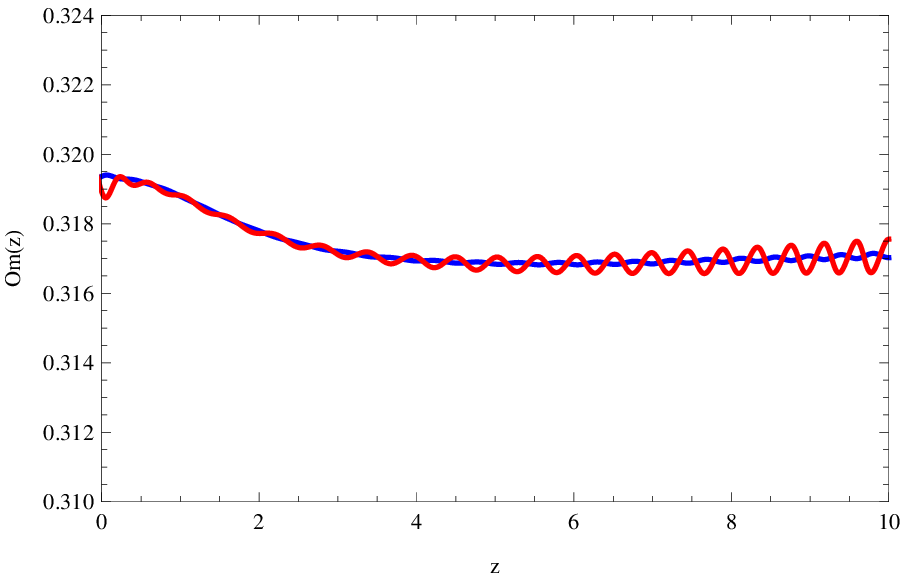}
\includegraphics[width=18pc]{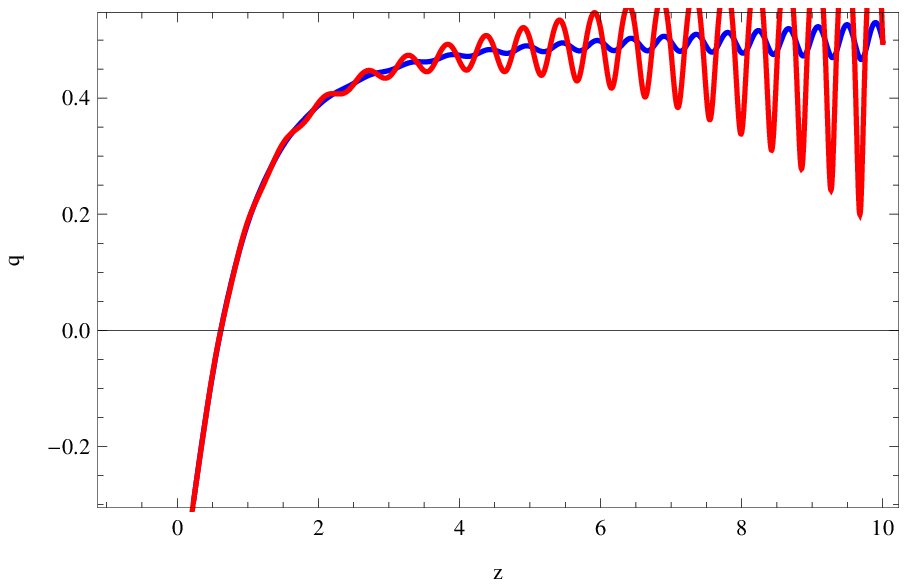}
\caption{The deceleration parameter $q$ (upper left plot), the
statefinder $Om(z)$ (upper right plot), and the statefinder
function $y_H$ (lower left) as functions of the redshift, for
$\tilde{\gamma}=1/10$ (red curves) and for $\tilde{\gamma}=1/10^3$
(blue curves).} \label{plot4}
\end{figure}

The function $y_H$ is by itself a statefinder parameter for the
dark energy era, and in fact was used indirectly in Ref.
\cite{Sahni:2014ooa} to discuss the issue that it might become
negative at a redshift $z\sim 2.34$ firstly reported in Ref.
\cite{Delubac:2014aqe}. Basically, good statefinder parameters are
those associated with the geometry of the spacetime, thus the
Hubble rate and its higher derivatives. Also it is known that the
$F(R)$ gravity models that can describe a dynamical dark energy
era, are plagued with the problem of dark energy oscillations at
high redshift $z>6$, and singularities in the dark energy EoS
parameter may occur during the matter domination era. In our case,
the $R^2$ term cures these singularities, as it is already known
in the literature that such a term amends the singularities issue
in the dark energy EoS \cite{Appleby:2009uf,Bamba:2008ut}, but the
dark energy oscillations issue still remains. In fact, as we will
show, the oscillations issue becomes more evident in statefinder
parameters that contain higher derivatives of the Hubble rate, and
is strongly affected by the initial conditions chosen for $y_H$,
thus it is affected by the parameter $\tilde{\gamma}$ appearing in
Eqs. (\ref{generalinitialconditions}). Let us solve numerically
the differential equation (\ref{differentialequationmain}), for
the values of the parameters defined in this and the previous
sections, with the initial conditions
(\ref{generalinitialconditions}), for $\tilde{\gamma}=1/10^3$. In
the left plot of Fig. \ref{plot1} we present the behavior of the
function $y_H$ as a function of the redshift, and already the
issue of dark energy oscillations becomes apparent for redshifts
$z\sim 4$ and higher. In the right plot of Fig. \ref{plot1} we
present the behavior of the scalar curvature
(\ref{ricciscalarasfunctionofz}) as a function of the redshift.
In order to better understand the behavior of the model and assess
the viability of the model, we shall compare several statefinder
quantities for the model and directly compare these to the
$\Lambda$CDM model values. We start off with the dark energy EoS
parameter $\omega_G=\frac{P_G}{\rho_G}$, which can be expressed in
terms of the function $y_H$ as follows,
\begin{equation}\label{omegade}
\omega_G(z)=-1+\frac{1}{3}(z+1)\frac{1}{y_H(z)}\frac{d y_H(z)}{d
z}\, ,
\end{equation}
which is also a good statefinder since it depends on the geometry
via the dependence on the derivatives of the Hubble rate $H(z)$.
In the left plot of Fig. \ref{plot2} we present the behavior of
the dark energy EoS for $\tilde{\gamma}=1/10^3$. As it can be
seen, the EoS parameter for the model has oscillating behavior for
$z\succeq 6$ approximately, but the present day value is
$\omega_G(0)=-0.995175$, which is compatible with the latest
Planck constraints \cite{Aghanim:2018eyx}, which is
$\omega_G=-1.018\pm 0.031$. The results for present day values of
the various parameters that will be obtained hereafter, are
summarized in Table \ref{table1}. Also in the right plot of Fig.
\ref{plot2} we plot $\Omega_G=\frac{\rho_G}{\rho_{tot}}$, as a
function of the redshift, and $\Omega_G(z)$ is written as a
function of the function $y_H(z)$ as follows,
\begin{equation}\label{omegaglarge}
\Omega_G(z)=\frac{y_H(z)}{y_H(z)+(z+1)^3+\chi (z+1)^4}\, .
\end{equation}
As it can be seen in the right plot of Fig. \ref{plot2}, the
late-time behavior of $\Omega_G(z)$ is oscillation-free and leads
to the present day value prediction $\Omega_G(0)=0.681369$, which
is also compatible with the latest Planck constraint
$\Omega_G=0.6847\pm 0.0073$.
The statefinder quantities as we already mentioned, are quite
important since the results depict the effects of the geometry of
spacetime on the statefinder quantities, and this is why the
statefinder quantities are valuable tools for late-time cosmology.
We shall be interested in four statefinder quantities, namely the
deceleration parameter $q$, the jerk $j$, the parameter $s$
\cite{Sahni:2002fz} and finally the parameter $Om(z)$
\cite{Sahni:2014ooa}, which their functional behavior as functions
of the Hubble rate is given below,
\begin{align}\label{statefinders}
& q=-1-\frac{\dot{H}}{H^2}\, , \,\,\,
j=\frac{\ddot{H}}{H^3}-3q-2\, , \\ \notag &
s=\frac{j-1}{3(q-\frac{1}{2})}\, , \,\,\,
Om(z)=\frac{\frac{H(z)^2}{H_0^2}-1}{(1+z)^3-1}\, .
\end{align}
The last three, namely the jerk $j$, $s$ and $Om(z)$ have very
simple values for the $\Lambda$CDM model, which are $s=0$, $j=1$
and $Om(z)=\Omega_M\simeq 0.3153$. We can easily compare the
results of the axion $F(R)$ gravity model with those of the
$\Lambda$CDM by expressing the statefinders (\ref{statefinders})
as functions of the redshift, but the final expressions are too
lengthy to quote here. The results of the numerical integration
are shown in Figs. \ref{plot3}, where in the upper left the
deceleration parameter of the axion $F(R)$ gravity model is
plotted as a function of the redshift (blue curve) and the
corresponding deceleration parameter of the $\Lambda$CDM model
also appears (red). As it can be seen, these are indistinguishable
up to a redshift $z\sim 4$. In the upper right of Fig. \ref{plot3}
, in the lower left we plot the jerk $j$ and in the lower right
plot the statefinder $Om(z)$ for the axion $F(R)$ model and for
the $\Lambda$CDM is shown. In all the plots, the red curve
corresponds to the $\Lambda$CDM model, and the blue curve to the
axion $F(R)$ gravity model, for $\tilde{\gamma}=1/10^3$. As it can
be seen in the upper plots of Fig. \ref{plot3}, when lower
derivatives of the Hubble rate are invoked, or even simply of the
Hubble rate in the case of the statefinder $Om(z)$, the
oscillations of the dark energy at higher redshifts are not so
pronounced. However, it is notable that the $\Lambda$CDM and the
axion $F(R)$ gravity model can be distinguished even at low
redshifts when the statefinder $Om(z)$ is considered. From the
lower plots of Fig. \ref{plot3} it is evident that when higher
derivatives of the Hubble rate are invoked, the dark energy
oscillations are strongly pronounced even for redshift values
$z\sim 2$ and higher, as we already expected. However, the low
redshift behavior of the statefinders $j$ and $s$ are similar to
the ones corresponding to the $\Lambda$CDM model.
We need to note though, that although the dark energy oscillations
are apparent especially in more complex statefinder parameters,
the $R^2$ term in the $F(R)$ gravity of Eq. (\ref{starobinsky})
ensures the absence of singularities in the dark energy EoS
parameter, as is also noted in the literature
\cite{Appleby:2009uf}. In fact, this can also be seen in the left
plot of Fig. \ref{plot2}, since the oscillations of dark energy
are not so pronounced, in comparison to statefinders containing
higher derivatives of the Hubble rate. The results of the values
of the statefinders at present time and the comparison the the
values of the $\Lambda$CDM model can be found in Table
\ref{table1}.
\begin{table}[h!]
  \begin{center}
    \caption{\emph{\textbf{Values of Cosmological Parameters for Axion $F(R)$ Gravity Model and $\Lambda$CDM for initial conditions with $\tilde{\gamma}=1/10^3$.}}}
    \label{table1}
    \begin{tabular}{r|r|r}
     \hline
      \textbf{Cosmological Parameter} & \textbf{Axion $F(R)$ Gravity Value} & \textbf{Base $\Lambda$CDM or Planck 2018 Value} \\
           \hline
      $\Omega_{G}(0)$ & 0.683948 & $0.6847\pm 0.0073$ \\
      $\omega_G(0)$ & -0.995205 & $-1.018\pm 0.031$\\
      $Om(0.000000001)$ & 0.319364 & $0.3153\pm 0.007$\\
      $q(0)$ & -0.520954 & -0.535\\
      $j(0)$ & 1.00319 & 1\\
      $s(0)$ & -0.00104169 & 0\\
      \hline
    \end{tabular}
  \end{center}
\end{table}
Another important issue we need to discuss is the effect of the
initial conditions on the phenomenology of the axion $F(R)$
gravity model. We performed the numerical integration of the
differential equation (\ref{differentialequationmain}) with the
initial conditions (\ref{generalinitialconditions}) for two values
of $\tilde{\gamma}$, namely $\tilde{\gamma}=1/10$ and
$\tilde{\gamma}=1/1000$. The results are quite interesting since
the effect of the initial conditions on the values of the physical
parameters and on the statefinders at present time is minor, apart
from the ones that contain higher derivatives of the Hubble rate,
namely $j$ and $s$, but the changes are of the order
$\mathcal{O}(10^{-2})$. For a direct comparison we quote these on
Table \ref{table2}. However, the effect of the values of
$\tilde{\gamma}$ and in effect of the initial conditions on the
dark energy oscillations is mentionable, and in fact, as
$\tilde{\gamma}$ takes larger values, the oscillations are more
pronounced. This can be seen in Fig. \ref{plot4}, where we present
the behavior of the deceleration parameter $q$, of the statefinder
$Om(z)$ and of the function $y_H(z)$ as a function of the
redshift, for $\tilde{\gamma}=1/10$ (red curves) and for
$\tilde{\gamma}=1/1000$ (blue curves). Also for even smaller
values of $\tilde{\gamma}$, the oscillation behavior is even less
pronounced.
\begin{table}[h!]
  \begin{center}
    \caption{\emph{\textbf{Values of Cosmological Parameters for Axion $F(R)$ Gravity Model and $\Lambda$CDM for initial conditions with $\tilde{\gamma}=1/10$.}}}
    \label{table2}
    \begin{tabular}{r|r|r}
     \hline
      \textbf{Cosmological Parameter} & \textbf{Axion $F(R)$ Gravity Value} & \textbf{Base $\Lambda$CDM or Planck 2018 Value} \\
 \hline
      $\Omega_{G}(0)$ & 0.683968 & $0.6847\pm 0.0073$ \\
      $\omega_G(0)$ & -0.995827 & $-1.018\pm 0.031$\\
      $Om(0.000000001)$ & 0.318919 & $0.3153\pm 0.007$\\
      $q(0)$ & -0.521621 & -0.535\\
      $j(0)$ & 0.980965 & 1\\
      $s(0)$ & 0.00621087 & 0\\
      \hline
    \end{tabular}
  \end{center}
\end{table}
In conclusion, we demonstrated that the axion $F(R)$ gravity model
can mimic the $\Lambda$CDM model at late-times and can also
provide a viable late-time phenomenology compatible with the
Planck 2018 constraints on the cosmological parameters quantifying
the effects of the dark energy. However, the issue of dark energy
oscillations seems to be present, as was expected though. The dark
energy oscillations are strongly affected by the initial
conditions chosen for the function $y_H(z)$ and its derivative
during the last stages of the matter domination era, however we
will report soon on a mechanism that may eliminate completely the
dark energy oscillations from $F(R)$ gravity models.

\begin{figure}[h!]
\centering
\includegraphics[width=18pc]{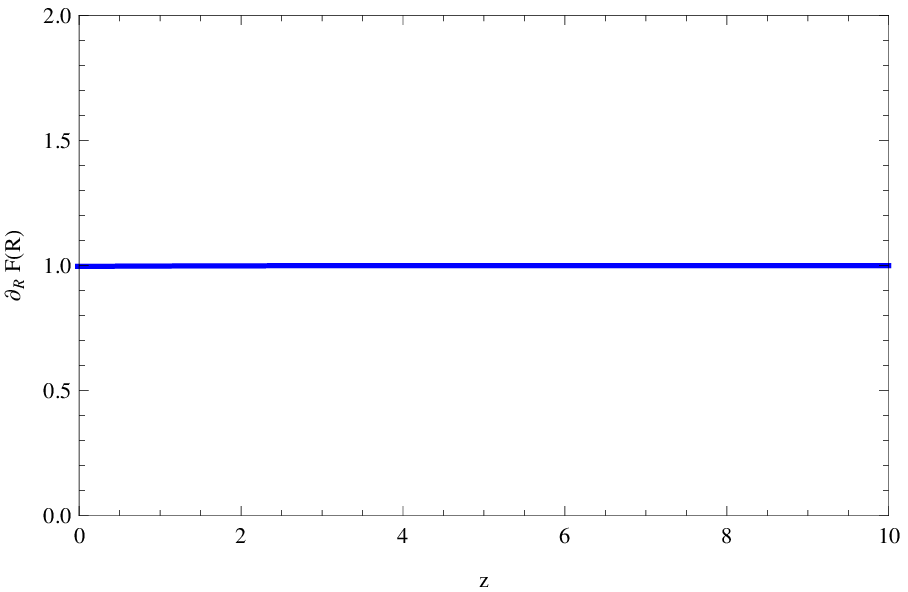}
\includegraphics[width=18pc]{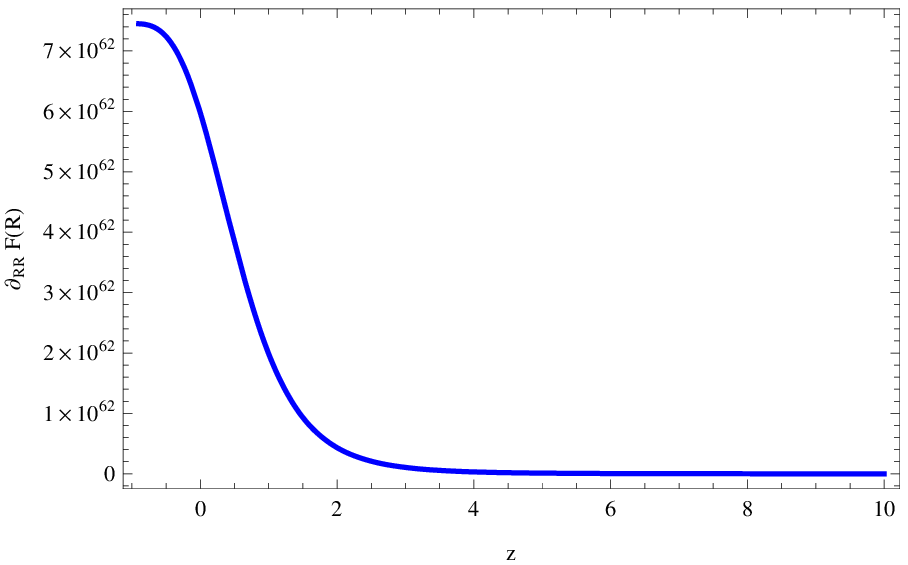}
\caption{ $F'(R)$ and $F''(R)$ as functions of the redshift. As it
can be seen, these are both positive thus the viability criteria
for the $F(R)$ function hold true.} \label{plot5}
\end{figure}

At this point, let us investigate whether the $F(R)$ gravity model
satisfies the viability criteria that any $F(R)$ gravity model
should satisfy. These are,
\begin{equation}\label{viabilitycriteria}
F'(R)>0\,, \,\,\,F''(R)>0\, ,
\end{equation}
for $R>R_0$, where $R_0$ is the present day curvature. In Fig.
\ref{plot5} we plot the behavior of $F'(R)$ and $F''(R)$ as
functions of the redshift, for small redshifts, and the same
applies for higher redshifts. In fact, when $H\sim H_I$, then
$R\sim 12H_I^2$ and we have approximately $F''(R)\sim 2.15\times
10^{-28}$eV$^{-1}$, and $F'(R)\sim 3.59$, thus the viability
conditions are satisfied even up to inflationary scales.

Before closing, we need to note that in Ref.
\cite{Odintsov:2017hbk} we also proposed an $F(R)$ gravity in
which the functional form of the $F(R)$ gravity was chosen in such
a way so that the dark energy oscillations in large redshifts are
reduced or eliminated. The action in the case of Ref.
\cite{Odintsov:2017hbk} was of the form,
\begin{equation}
I=\int_\mathcal{M}d^4\sqrt{-g}\left[\frac{R}{\kappa^2}+\gamma(R)R^2+
f_{\text{DE}}(R)+\mathcal L_m\right]\, ,\label{action}
\end{equation}
where,
\begin{equation}
f_\text{DE}(R)=-\frac{2\Lambda g(R)(1-\text{e}^{-b
R/\Lambda})}{\kappa^2}\,,\quad 0<b\,,\label{fDE}
\end{equation}
with $b$ being a positive parameter and $\Lambda$ being the
cosmological constant. In addition, the function $g(R)$ is,
\begin{equation}
g(R)=\left[1-c\left(\frac{R}{4\Lambda}\right)\log\left[\frac{R}{4\Lambda}\right]\right]\,,\quad
0<c\,,\label{gR}
\end{equation}
with $c$ being a real and positive. The choice of the function
$g(R)$ is crucial for the stabilization of the theory at large
redshifts, and it is basically a deformed $R^2$ correction. As was
shown in Ref. \cite{Odintsov:2017hbk}, such a stabilization is
achieved in this case, but we do not perform this analysis in this
paper.

\section{Concluding Remarks}

In this work we considered an axion $F(R)$ gravity model, in which
the axion eventually is the main component of cold dark matter in
the Universe, and we demonstrated that the axion $F(R)$ gravity
model can unify the early-time with the late-time acceleration. We
chose the $F(R)$ gravity to contain an $R^2$ term and also a term
$R^{\delta}$ with $0<\delta<1$, with the $R^2$ term being
motivated for a viable description of the inflationary era, while
the term $R^{\delta}$ was added for a successful description of
the late-time acceleration era. The axion scalar field was
described by a canonical scalar field with a broken primordial
$U(1)$ Peccei-Quinn symmetry, in the context of the misalignment
axion scenario. During the inflationary era, the axion scalar was
frozen in its primordial vacuum expectation value, and thus it
merely contributes a cosmological constant in the gravitational
equations of motion. In effect, the $R^2$ term of the $F(R)$
gravity controls the primordial dynamics, and it generates a
viable acceleration era. We quantified these considerations and we
demonstrated that indeed the axion and the $R^{\delta}$ terms have
a minor contribution to the dynamics of the model. As the Universe
expands though, when $H\sim m_a$ and for $m_a\gg H$, the axion
starts to oscillate. Assuming a slowly varying oscillation for the
axion, we demonstrated that the axion energy density scales as
$\rho_a\sim a^{-3}$, thus the axion scalar mimics the dark matter
fluid, with an average EoS parameter $w_a\sim 0$. At late times,
the $F(R)$ gravity term $R^{\delta}$ controls the dynamics of the
model, via its first derivatives, but we needed to quantify this
in the best way we could, so we used numerical analysis to solve
the Friedmann equation. We introduced a statefinder function
$y_H(z)$ which is the fraction of the dark energy energy density
over the current cold dark matter energy density
$y_H=\rho_G/\rho_m^{(0)}$, and we rewritten the Friedmann equation
in terms of $y_H$. By using appropriate initial conditions, we
numerically solved the Friedmann equation, and we focused our
analysis up to redshift $z_f\sim 10$. We focused our analysis on
the behavior of statefinder quantities, and also on the energy
density of dark energy and its EoS behavior. We found that the
$F(R)$ gravity model produces results very similar to the
$\Lambda$CDM model, in some cases almost identical for small
redshifts, and in all cases compatible results with the latest
Planck constraints on the cosmological parameters. In addition, we
showed that for statefinder quantities that contain higher
derivatives of the Hubble rate, the dark energy oscillations issue
occurs, as expected. This oscillation issue is more pronounced for
larger redshifts, and is strongly affected by the initial
conditions we used. However, the presence of the $R^2$ term
ensures the absence of singularities in the dark energy EoS
parameter, nevertheless this issue of oscillations, somewhat
obscures the whole picture. In response to this issue, in a future
work we shall present how the oscillations can be eliminated from
the $F(R)$ gravity late-time phenomenology.

Another issue we did not address is related to the reheating era.
In the axion $F(R)$ gravity model we present, this is expected to
be somewhat complicated, and not so easy to address analytically,
due to the presence of $a^{-3}$ and $R^{\delta}$ terms caused by
the axion and the $F(R)$ gravity in the Friedmann equation. In
principle, the $R^2$ term is not expected to be dominant in this
era, since the curvature is already too small, and the axion
already starts to oscillate in a slowly-varying way. This study is
an important one, which we hope to address in a future work
focused on this issue.

Another issue that is quite interesting and valuable for future
observations, of matter curvature perturbations for the $F(R)$
gravity axion model. In the presence of $F(R)$ gravity, Newton's
effective gravitational constant is different from the present day
value, during the matter domination era, however, the axion also
contributes to the matter curvature perturbations. The question is
which contribution is dominant and to which extent. The study for
the axion effect on the matter curvature perturbations was
performed in
\cite{Choi:1999zy,Hwang:2009js,Park:2012ru,Noh:2013coa,Noh:2015spa},
while for the $F(R)$ gravity case see \cite{Bamba:2012qi}. The
evolution of the matter curvature perturbations is valuable
phenomenologically, since the growth index is an observable that
may even discriminate modified gravities between them.

A highly non-trivial issue is the axion isocurvature perturbations
issue, generated during or mainly well after the first horizon
crossing and before the horizon re-entry. Although misalignment
axion scalars have minor backreaction from isocurvature
perturbations during inflation, if these are generated well after
the first horizon crossing, might have an observable effect. The
question is to what extent may $F(R)$ gravity affect the
generation of axion isocurvature perturbations and their
evolution. This is a highly non-trivial issue to address, and we
leave this open as question.

Finally, it is noteworthy another phenomenologically interesting
issue, which deserves to be addressed in detail in a future work.
It is related to the fact that in some observational data
\cite{Sahni:2014ooa,Delubac:2014aqe}, these suggest that the
function $y_H\sim \rho_G$ took negative values during the last
stages of the matter domination era, and specifically around
$z\sim 2.34$. This is a highly intriguing issue, and we devised a
mechanism in the context of modified gravity in order to produce
both a viable late-time era at $z\sim 0$ and a negative $y_H$,
without introducing compensating dark energy mechanisms. We shall
report on this intriguing issue in the near future.

\section*{Acknowledgments}

This work is supported by MINECO (Spain), FIS2016-76363-P, and by
project 2017 SGR247 (AGAUR, Catalonia) (S.D.O), and by Russian
Ministry of Science and High Education, project No. 3.1386.2017.

\end{document}